\documentclass[superscriptaddress,nobibnotes,prb,twocolumn]{revtex4-1}

\usepackage{amsmath,amssymb} 
\usepackage[normalem]{ulem}
\usepackage{braket}
\usepackage{dsfont}
\usepackage{graphicx}
\usepackage{dcolumn}
\usepackage{bm}
\usepackage{color}
\usepackage{lipsum}
\usepackage{siunitx}
\usepackage{hyperref}
\hypersetup{
    colorlinks=true,
    linkcolor=blue,
    filecolor=magenta,      
    urlcolor=blue,
}
\usepackage{multirow}
\newcommand{\hc}[1]{{#1}^{\dagger}}

\usepackage[dvipsnames]{xcolor}

\begin{document}

\title{Exciton dissociation mediated by phonons in organic photovoltaics}

\author{Stepan Fomichev}
\thanks{These authors contributed equally.}
\affiliation{Department of Physics and Astronomy, University of British Columbia, Vancouver, British Columbia, V6T 1Z1 Canada}
\affiliation{Stewart Blusson Quantum Matter Institute, University of British Columbia, Vancouver, British Columbia, V6T 1Z4 Canada}

\author{Leonard Ruocco}
\thanks{These authors contributed equally.}
\affiliation{Department of Physics and Astronomy, University of British Columbia, Vancouver, British Columbia, V6T 1Z1 Canada}
\affiliation{Stewart Blusson Quantum Matter Institute, University of British Columbia, Vancouver, British Columbia, V6T 1Z4 Canada}

\author{Alexandra Tully}
\affiliation{Department of Physics and Astronomy, University of British Columbia, Vancouver, British Columbia, V6T 1Z1 Canada}
\affiliation{Stewart Blusson Quantum Matter Institute, University of British Columbia, Vancouver, British Columbia, V6T 1Z4 Canada}

\author{Mona Berciu}
\affiliation{Department of Physics and Astronomy, University of British Columbia, Vancouver, British Columbia, V6T 1Z1 Canada}
\affiliation{Stewart Blusson Quantum Matter Institute, University of British Columbia, Vancouver, British Columbia, V6T 1Z4 Canada}
\affiliation{Leibniz Institute for Solid State and Materials Research (IFW) Dresden, Helmholtzstrasse 20, 01069 Dresden, Germany}

\date{\today}

\begin{abstract}
It is well known that phonons can overscreen the bare Coulomb electron-electron repulsion, turning it into the effective attraction that binds the Cooper pairs responsible for  BCS superconductivity. Here, we use a simple lattice model to prove that the counterpart of this is also possible, whereby phonons overscreen the bare electron-hole attraction and may turn it repulsive at short distances, driving exciton dissociation in certain regions of the parameter space. We argue that this phonon-mediated short-range screening plays an important role in the physics of organic solar cell materials (and other materials with strong electron-phonon coupling) and could point the way to new strategies for optimizing their efficiencies.
\end{abstract}
\maketitle
\noindent

\section{Introduction}
Organic solar cells (OSCs) have been heralded as a revolutionary technology in the renewable energy sector due to their flexible and light-weight nature and low production cost.\cite{Kaltenbrunner:2012, Xu:2018, Gambhir:2016, chen19} While power conversion efficiencies of OSC devices have been improving,\cite{dosSantosRosa:2021} they have not yet reached levels high enough for OSCs to realize their promise; this is largely due to the challenge of efficiently extracting free charge carriers without detrimental losses.\cite{Heeger:2013, Vandewal:2020} 

All light-harvesting devices start by capturing a photon to excite a bound electron-hole pair -- an exciton. {Voltage} is ultimately produced through the generation of free charge carriers, requiring the dissociation of the exciton through some internal mechanism.

Conventional (inorganic) solar cells, such as those based on Si or GaAs, have highly effective charge screening. Because the screened Coulomb attraction is  weak, the Wannier excitons it creates are highly extended and have small binding energies (few tens of meV). A combination of thermal fluctuations and external electric fields is therefore sufficient to drive dissociation. 

By contrast, OSC materials have poor charge screening, resulting in small Frenkel excitons with large binding energies of a hundred meV or more.\cite{Gledhill:2005,Nelson:2002} These are stable against thermal fluctuations and fairly long-lived, leading to high recombination losses and reduced efficiencies. This is why understanding and engineering exciton dissociation in OSCs remains a foundational challenge.

To date, the most investigated approach to engineering dissociation is to use bulk-heterojunction interfaces combining donor and acceptor materials, chosen so that the potential gradient at their interface helps overcome the high binding energies. This setup was shown to produce higher yields, which was attributed to enhanced dissociation of so-called charge-transfer states at the donor/acceptor (D/A) interface.\cite{gelinas14,Sutty:2014} Charge-transfer states are believed to be relatively short-lived excitons composed of an electron and a hole that span neighbouring molecular sites. While such excitons are quite commonly generated in the bulk,\cite{emmerich2020} they delocalize more easily when they span a D/A interface.\cite{Bakulin:2012, Bernardo:2014,Kahle:2018} However, more work is needed to understand both the nature of these states, and how they can be engineered to optimize exciton dissociation.

Alongside charge screening, the vibrational characteristics (phonon modes) of OSCs are also relevant to dissociation -- and even less well-understood. Phonons couple strongly to molecular orbitals, as evidenced by photoemission experiments,\cite{canton02} and thus may be playing a role in the exciton dynamics.\cite{Zhugayevych:2015} Most of the studies to date have focused on the role of phonons in the formation of charge transfer states,\cite{falke14,song14} and how electron-phonon coupling affects the yield across the D/A interface.\cite{bera15,hu21,jailaubekov13,tamura13,tamura08,bittner14}

\begin{figure}[t]
\centering
\includegraphics[width=0.45\textwidth]{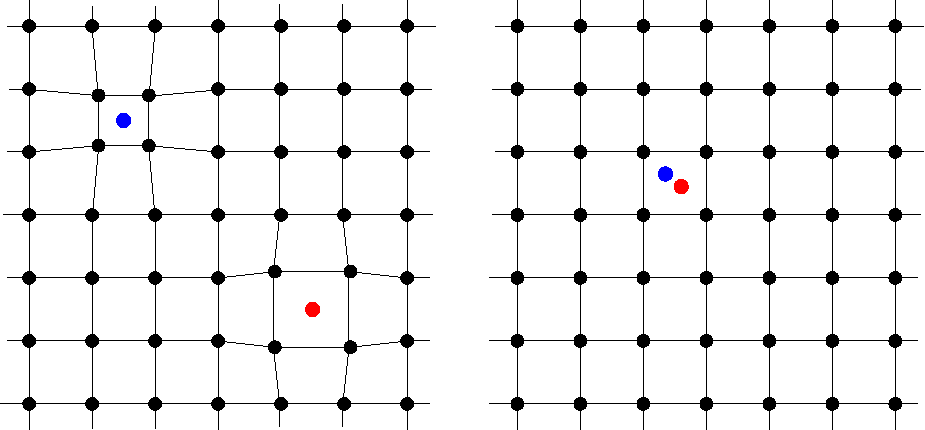}
				\caption{ Lattice distortion from an exciton. Left panel: when the electron and the hole are far apart (red and blue circles, respectively) their excess charge induces local lattice distortions, giving rise to polarons. Right panel: A small Frenkel exciton produces a much weaker electric potential and thus a much smaller lattice distortion. }
		\label{fig:excdiag}
\end{figure}

Here we present a fundamentally different way whereby electron-phonon coupling can influence exciton dissociation, even in the absence of a D/A interface.  We show that sufficiently strong electron-phonon coupling can be \textit{directly responsible} for exciton dissociation, despite the presence of significant Coulomb attraction between the electron and the hole. 

The basic idea is sketched out in Fig. \ref{fig:excdiag}, where we compare the effects of electron-phonon coupling when the hole and electron are far apart (left panel) versus when bound in a small exciton (right panel). The addition of an excess carrier results in a local lattice distortion that dresses that carrier into a polaron. Because the electron and the hole have opposite charges, in a polar material they create opposite lattice distortions in their vicinity. However, when they are bound into a small exciton, their clouds essentially cancel each other out, and locally there is no distortion. Another way to say this is that there is no excess local charge in the presence of a small exciton -- hence no local lattice distortion is expected.  

In this picture, electron-phonon coupling is seen to lower the energy of the dissociated state through polaron formation, while having little effect on the exciton binding energy. For large enough electron-phonon coupling this leads to outright dissociation, as we show next. Even when that is not the case, our work shows that one must take polaron formation into consideration when choosing the donor/acceptor materials, because the polaronic contribution to the energetic landscape can be considerable.  


It is important to acknowledge that the idea of exciton dissociation driven by electron-phonon coupling was proposed previously by 
Sumi in Ref. \onlinecite{sumi77}, where he used a variational approximation to study the effect of Fr\"ohlich coupling on an exciton. His prediction of a sharp transition between bound (exciton) and unbound (free electron and hole polarons) states was later discredited by Gerlach and L\"owen,\cite{gerlach90} who proved that sharp transitions are forbidden in this class of Hamiltonians and concluded that overscreening is impossible in this context. We find a smooth crossover between the two types of states, fully consistent with the mathematical proof of Ref. \onlinecite{gerlach90}. Our work shows that the contradiction between Refs. \onlinecite{sumi77} and \onlinecite{gerlach90} is not because overscreening is impossible, but because the predicted sharp transition was an artifact of the variational approximation\cite{pollmann77} used by Sumi. 

The article is organized as follows: Sec. \ref{sec:model} introduces the model we use to study this problem, and Sec. \ref{sec:methods} explains our formalism and approach. Key results are shown in Sec. \ref{sec:results}, while Sec. \ref{sec:discussion} contains an extended discussion of the various approximations made in the model and the relevance of this phenomenology in the context of OSCs.

\section{The Model}
\label{sec:model}

We consider a single electron-hole pair in a one-dimensional (1D) ionic chain, where each site supports a single on-site orbital and a  dispersionless Einstein phonon mode. The single electron-hole pair assumption is reasonable if, for example, the concentration of photo-generated electron-hole pairs in the material is very low. We focus on the 1D chain because here it is known that Coulomb attraction always results in the formation of strongly bound excitons, unlike in higher dimensions where excitons can be either exponentially weakly bound (in 2D) or unstable unless the attraction is sufficiently strong (in 3D).\cite{burovski08} Thus, demonstrating dissociation in 1D would imply similar behaviour in higher dimensions, given that the exciton is even more loosely bound there.

Our Hamiltonian reads:
\begin{equation}
\hat{H} = \hat{T}_e +\hat{V}_{e-h} + \hat{H}_{ph} + \hat{V}_{e-ph}
+\hat{V}_{h-ph}.
\label{eq:ham_exc_diss}
\end{equation}

Here, $\hat{T}_e=\sum_{k\sigma} \epsilon_{k} c_{k\sigma}^{\dagger}c_{k\sigma}$ is the kinetic energy of free electrons in the conduction band, described by a tight-binding model with a dispersion $\epsilon_k = -2t\cos k$ defined by the hopping $t$ and momentum $k\in (-\pi,\pi]$ of the bare electron (the lattice constant is set to $a=1$, also $\hbar=1$). The creation operator $c_{k\sigma}^{\dagger}$ adds an electron with momentum $k$ and spin $\sigma$ in this band. Its real space counterpart is $c_{n\sigma}^{\dagger}$, where $n=1\dots N$ indexes the sites of the  chain, with $N\rightarrow \infty$.
 Hole creation operators in real space are denoted by $h_{n\sigma}^{\dagger}$. For simplicity, we assume that holes are localized (we reflect on this assumption in Sec. \ref{sec:discussion}).

The electron-hole interaction $\hat{V}_{e-h}$ is modeled as an on-site Coulomb attraction 
\begin{equation}
\hat{V}_{e-h}=- U \sum_{n,\sigma,\sigma'} h_{n\sigma}^{\dagger}h_{n\sigma} c^{\dagger}_{n\sigma'}c_{n\sigma'},
\end{equation}
characterized by $U>0$. Longer (but finite) range attractions can be treated similarly and lead to quantitative changes only, at the cost of adding more parameters.

Optical phonons are described with an Einstein model:
$$
\hat{H}_{ph}=\Omega\sum_n b_n^{\dagger}b_n \\
$$
where $b^\dagger_n$ creates a phonon with energy $\Omega$ at site $n$. 

Finally, the Holstein carrier-lattice couplings are:
\begin{align}
\hat{V}_{e-ph}=& M_e \sum_{n\sigma}c_{n\sigma}^{\dagger}c_{n\sigma} (b_n+b_n^{\dagger}) \\
\hat{V}_{h-ph}=& M_h \sum_{n\sigma}h_{n\sigma}^{\dagger}h_{n\sigma} (b_n+b_n^{\dagger}) 
\end{align}
with electron/hole-phonon couplings $M_e$ and $M_h$, respectively. 

Even after all these simplifications, there are four dimensionless parameters: $U/t, \Omega/t, M_e/t, M_h/t$. To avoid further complications, we set the temperature $T=0$. This is justified because we are interested in cases where all energy scales (including the exciton binding energy) are much larger than the thermal energy, as is typically the case in organic photovoltaics. 

\section{Methods}
\label{sec:methods}
Finite Coulomb attraction in 1D always leads to a ground-state with a stable, bound exciton. Our aim is to investigate the influence of the carrier-phonon couplings on the stability of the exciton. To do this, we calculate the  Green's function
\begin{equation}
G_{ij}(z) \equiv \bra{0} c_i h_i \hat{G}(z)  h_i^{\dagger}c_j^{\dagger}\ket{0}
\label{eq:Greens_exciton}
\end{equation}
where we reserve the index $i$  to label the site hosting the immobile hole (the spin degree of freedom is irrelevant for this calculation and we ignore them from now). The electron can move and the propagator above is the Fourier transform (at energy $z=\omega + i \eta$) of the amplitude of probability that if the hole is at site $i$, the electron moves from site $j$ to site $i$ within a given time interval, with both the initial and the final states having no phonons $b_n|0\rangle =0$. The broadening $\eta \rightarrow 0$ introduces an artificial lifetime $\propto 1/\eta$ for the pair to recombine, and $\hat{G}(z)=(z-\hat{H})^{-1} $ is the resolvent. The associated local density of states (LDOS), plotted in the figures, is defined as $A(\omega)  = -\mathrm{Im} G_{ii}(z) / \pi$;  invariance to translations ensures that the LDOS is the same at all sites $i$.

The propagator of Eq. (\ref{eq:Greens_exciton}) for the full interacting Hamiltonian is calculated using a novel, generalized version of the Momentum Average approximation (MA) -- a method well established and validated for studying single polarons\cite{berciu06,berciu07,Berciu2010,marchand10} and bipolarons.\cite{adolphs14,sous17,sous18} This generalization allows, for the first time, to include into the variational space configurations with two phonon clouds located arbitrarily far apart: a hole cloud at site $i$, and an electron cloud elsewhere in the chain. 

We now briefly describe this method, before moving to discuss the results.

\subsection{Non-interacting spectrum}
\label{sec:methods_nonint}
The first step is to obtain the Green's function in the absence of carrier-phonon coupling ($M_e = M_h = 0$). The Green's function $G_{ij}^{(i,0)}(z)$ corresponding to  $\hat{H}_0 = \hat{T}_e +\hat{V}_{e-h} +\hat{H}_{ph}$, {\it i.e.} for the system without carrier-phonon coupling, can be calculated analytically (see Appendix \ref{app:impurity_Greens_funs} for details). The spectrum extracted from the poles of this Green's function has a discrete eigenstate at $\omega=-\sqrt{4t^2+U^2}$ and a continuum for $\omega \in [-2t, 2t]$. The continuum describes the electron unbound to the hole, {\em i.e.}  free to move throughout the system. The discrete eigenstate is the energy of the bound exciton, lying below this continuum for any value of $U>0$. All these features would be shifted by $n\Omega$ if there were $n$ phonons in the system, but for the propagator of interest to us $n=0$.

\subsection{Turning on interactions: Lang-Firsov transformation}

In the presence of carrier-phonon couplings (finite $M_e, M_h$), if the  carriers are not bound then they each create phonon clouds in their vicinity, turning into polarons. In the bound state their clouds combine, resulting in an exciton-polaron.

Because the hole cannot move in our simplified model, and because its coupling to the lattice is local, its phonon cloud is definitely located at hole site $i$. 
We then use the Lang-Firsov transformation $\mathcal{U}_i =\exp[ \frac{M_f}{\Omega} (b_i - b_i^{\dagger})]$  to integrate out the hole-phonon coupling:
\begin{multline}
\tilde{H}_i =  \mathcal{U}^\dagger_i\hat{H} \mathcal{U}_i =\hat{T}_e - \left( U + \frac{2 M_e M_h}{\Omega} \right) c^{\dagger}_i c_i - \frac{M_h^2}{\Omega} +\\
+ \Omega\sum_lb_l^{\dagger}b_l  + M_e \sum_l c_l^{\dagger}c_l (b_l^{\dagger}+b_l)
\end{multline}
 after noting that $ \mathcal{U}^\dagger_ib_l \mathcal{U}_i= b_l - \delta_{i,l}\frac{M_h}{\Omega}  $.
This transformation is exact and shows the hole-polaron formation energy $-M_h^2/\Omega$ but also a change of the effective Coulomb attraction experienced by the electron when at site $i$, $U\rightarrow \tilde{U}= U+\frac{2 M_e M_h}{\Omega}$,  arising from the electron's coupling to the hole's cloud in addition to the Coulomb interaction with the hole. The propagator for the electron-hole pair
\begin{equation}
G_{ij}(z) \equiv \bra{0} c_i h_i \hat{G}(z)  h_i^{\dagger}c_j^{\dagger}\ket{0}
\label{eq:Greens_exciton_app}
\end{equation}
is then rewritten in terms of the transformed Hamiltonian:
\begin{equation}
{G}_{ij}(z) = e^{-\frac{M_h^2}{2\Omega^2}}\sum_{n=0}^{\infty}\frac{1}{n!}\left(\frac{M_h}{\Omega}\right)^n H_{ij}(n,\tilde{z})
\label{eq:coup_exc_prop}
\end{equation}
where the new propagators
\begin{equation}
H_{ij}(n,\tilde{z}) = \bra{0}h_i c_i \mathcal{U}_i \tilde{G}(\tilde{z})h_i^\dagger c_j^{\dagger}b_i^{\dagger n}\ket{0}
\label{eq:H_ij_def}
\end{equation}
describe the propagation of the electron in the presence of phonons created by the hole. 
To obtain Eq. (\ref{eq:coup_exc_prop}) we used the Baker–Campbell–Hausdorff formula to rewrite $\mathcal{U}_i^\dagger\ket{0} = e^{-M_h^2/2\Omega^2}\sum_{n=0}^{\infty}\frac{1}{n!}\left(b_i^{\dagger} \frac{M_h}{\Omega}\right)^n \ket{0}$, and we introduced $\tilde{z} = z+M_h^2 / \Omega$ and the transformed resolvent
\[
 \mathcal{U}^\dagger_i\hat{G}(z) \mathcal{U}_i\equiv \tilde{G}(\tilde{z}) =(\tilde{z} - \hat{h}_i )^{-1} 
\]
where 
\[
 \hat{h}_i = \tilde{H}_i + \frac{M_h^2}{\Omega} = \hat{T}_e - \tilde{U} c^\dagger_i c_i + \hat{H}_{ph}+\hat{V}_{e-ph}
\]
describes the electron's kinetic energy, effective interaction with the hole located at $i$, and coupling to the lattice.
So far, everything is exact.
\subsection{Analogy to the disorder MA}

Note that $\hat{h}_i$ obtained above is formally equivalent to the Hamiltonian for an electron with Holstein coupling in the presence of an on-site `disorder' at site $i$.  In previous work, we have already demonstrated that for such problems, even the simplest version of the variational momentum average (MA) approximation, namely the one-site MA$^{(0)}$ version, is quantitatively accurate if $t/\Omega$ is not too large.\cite{MonaEPL89(2010), HadiPRB85(2012)} We use the same approximation here, straightforwardly generalized to include the presence of phonons created by the hole at site $i$. Specifically, we implement an MA where the variational space allows for the presence of two phonon clouds: one at site $i$ due primarily to the hole, and one at any other site of the system, created by the electron. We note that the electron cloud can be allowed to spread over more sites,\cite{DominicPRB95(2017)} increasing the accuracy of the approximation: however, the resulting improvements are quantitatively small and do not affect the physics. For our purposes it suffices to proceed with the one-site cloud approximation, which predicts energies to within an accuracy of a few percent.\cite{berciu06,GlenPRB74(2006),MonaEPL89(2010), HadiPRB85(2012)}

Proceeding by analogy with the disorder MA calculation, the equations-of-motion (EOMs) for the propagators in this two-cloud generalization of MA are obtained by repeated use of the Dyson identity $\hat{G} = \hat{G}_0 + \hat{G} \hat{V} \hat{G}_0$ with $\hat{V} = \hat{V}_{e-ph}$. The resulting system of equations (\ref{eq:H_ij_expand_dyson}-\ref{eq:f_system}) and its derivation are shown for completeness in Appendix \ref{app:inter_Greens_funs}. This linear system that emerges turns out to be amenable to further simplifications driven by the intuition that not all propagators contribute equally: indeed, we find that about half the propagators may be set to zero (halving the size of the system) with no noticeable changes to the resulting spectrum. More details on this further approximation and the intuition behind it are given in  \ref{app:simplify}, and in Appendix \ref{app:approx_full_compare} we show some results that justify the validity of this futher approximation.

\subsection{Exciton wavefunction and the phonon cloud} 
\label{sec:methods_char}

Once the Green's functions $G_{ij}$ are obtained by solving the linear system, to further elucidate the nature of the ground-state properties of our model we characterize the spatial extent of the exciton wavefunction, as well as calculate the size of its phonon cloud. To obtain the former, we use the Lehmann decomposition $G_{ij}({z}) = \sum_n { \braket{0 | h_ic_i | \psi_n} \braket{\psi_n | \hc{c}_j \hc{h}_i| 0} }/(z - E_n)$, where  $\hat{H}|\psi_n\rangle = E_n |\psi_n\rangle$ are the eigenstates with one electron and one hole. At the exciton energy $E_0$, and if $\eta$ is much smaller than the gap to the continuum, there is only one dominant contribution to the Lehmann sum: $G_{ij}({z= E_0+i\eta}) \approx { \braket{0 | h_i c_i | \psi_0} \braket{\psi_0 | \hc{c}_j \hc{h}_i| 0} }/i\eta $. Therefore we can use
\begin{equation}
    \rho_{ij}(E_0) = \frac{\left| \braket{0 | h_i c_j | \psi_0 } \right|^2 }{\left| \braket{0 | h_i c_i | \psi_0 } \right|^2}\approx  \frac{|G_{ij}(E_0)|^2}{|G_{ii}(E_0)|^2}
    \label{eq:rho}
\end{equation}
to characterize the probability that the electron is at a distance $|j-i|$ from the hole in the exciton ground-state, scaled such that $\rho_{ii}(E_0) =1$. 

\begin{figure*}
\centering
        \includegraphics[width=0.45\textwidth]{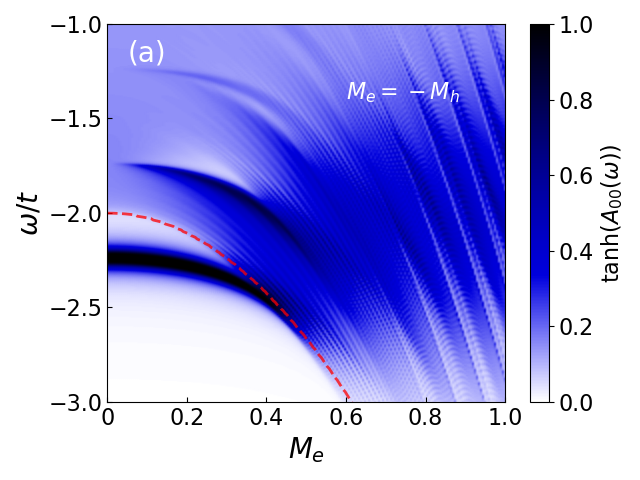}
        \includegraphics[width=0.45\textwidth]{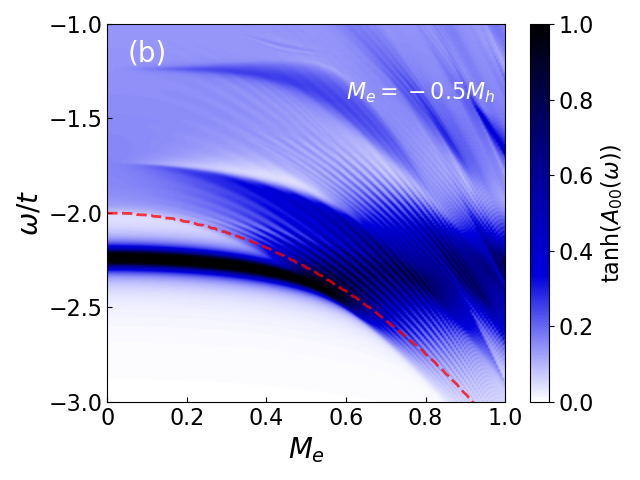}
        \includegraphics[width=0.45\textwidth]{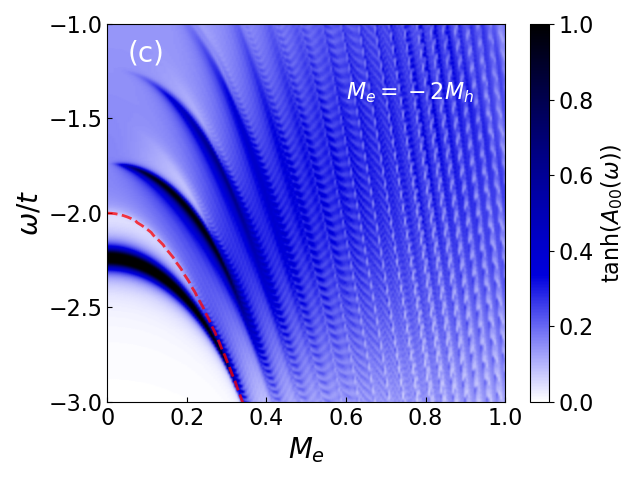}
        \includegraphics[width=0.45\textwidth]{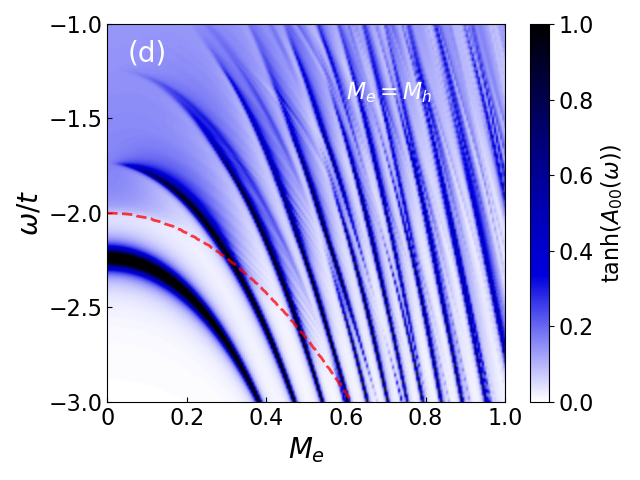}
\caption{Contour plots of the LDOS $A(\omega)$ at the hole site when $\Omega=0.5$ and $U=1$. The electron-phonon coupling $M_e$ is shown on the $x$ axis (the corresponding $M_h$ is indicated on the figure).  The dashed red line shows where we expect the lower edge of the continuum of eigenstates describing unbound electron- and hole-polarons, based on their  individually calculated MA energies. Its good agreement with the calculated spectral weight provides a validation of  the generalized MA we developed. The fast oscillations in the continuum weight are finite size effects, due to the cutoff $|l-i|_m=50$ for the maximum distance between the two clouds; the maximum numbers of phonons in the two clouds are set to $n_m=k_m=20$, sufficient for convergence.  The discrete peak appearing below the continuum at small $M_e$ is the exciton bound state, broadened into a Lorentzian by the finite $\eta=0.01$. With increasing coupling, the exciton approaches the continuum and eventually merges smoothly with it, marking its dissociation into a pair of unbound electron and hole polarons. This behaviour is robust so long as the couplings are of opposite sign, so that $M_e M_h <0 $, see panels (a)-(c). In contrast, when $M_e M_h>0$, the exciton is always stable, see panel (d).}
\label{fig:ldos_coupling_diags}
\end{figure*}

To calculate the average number of phonons $N_{\text{ph}}$ in the exciton cloud, we use the Hellmann-Feynman theorem:\cite{hellmann1937,feynman1939}
\begin{equation}
    N_{\text{ph}} =  \braket{ \psi_0 | \sum_l\hc{b}_l b_l | \psi_0  } = \frac{\partial E_0}{\partial \Omega}.
    \label{eq:num_ph_formula}
\end{equation}
The derivative is computed numerically with the  finite-difference approach. Both of these metrics give additional glimpses at the impact of phonons on the dissociation process.

\section{Results}
\label{sec:results}
\subsection{Exciton dissociation driven by electron-phonon coupling}
Armed with the methods from the previous section, we calculate the spectrum of a system with one electron and one hole, in the presence of short range (on-site) Coulomb attraction of magnitude $U>0$, and of Holstein carrier-phonon couplings $M_e$ and $M_h$, respectively, to an optical dispersionless phonon mode of energy $\Omega$. As stated previously, we focus on 1D chains, where the carriers' tendency to bind into an exciton is enhanced. The electron's nearest neighbor hopping is $t=1$; meanwhile the hole is localized, modeling either a valence band with a very large effective mass or a hole trapped by an acceptor impurity. 

Exciton dissociation driven by the electron-phonon coupling is demonstrated graphically in Fig. \ref{fig:ldos_coupling_diags}. The panels show the contour plot of the LDOS $A(\omega)$ at the hole site versus energy and coupling $M_e$, when $U=1$, $\Omega=0.5$ and $M_e=-M_h$ (panel a); $M_e=-0.5M_h$ (panel b); $M_e=-2M_h$ (panel c); and $M_e=M_h$ (panel d).

At $M_e=M_h=0$, the lowest energy feature in the electron+hole spectrum is a discrete peak marking the existence of the exciton, just as discussed in Sec. \ref{sec:methods_nonint}. If $M_eM_h<0$,  the discrete peak merges smoothly with the continuum at $M_e^{(c)}$ and  {\it the exciton dissociates into unbound electron- and hole-polarons} for $M_e > M_e^{(c)}$. There is no discontinuity in the LDOS at $M_e^{(c)}$: thus, there is no contradiction between our result and Ref. \onlinecite{gerlach90}. By contrast, if $M_eM_h>0$ (panel d), the exciton is further stabilized by increasing coupling.\cite{burovski08}

The carrier-phonon coupling $M$ is set by the gradient of the carrier-lattice potential with respect to a small lattice displacement. Because the hole and the electron have opposite charge, their respective carrier-lattice potentials have opposite signs and thus $M_e$ and $M_h$  have opposite signs. Physically, this is because a lattice distortion that is energetically favorable for an electron is generically unfavorable for a hole (left panel of Fig. \ref{fig:excdiag}).
Moreover, a very small Frenkel exciton, with the electron and hole at the same site,  creates no local charge imbalance so no lattice distortion is expected (right panel of Fig. \ref{fig:excdiag}). In the atomic limit ($t=0$), a vanishing exciton-polaron binding energy   $-(M_e+M_h)^2/\Omega\approx 0$  implies that $M_e\approx -M_h$. Of course, one can envision more complex situations where $|M_e|\ne|M_h|$, however panels (b) and (c) of Fig.~ \ref{fig:ldos_coupling_diags} show the same  dissociation phenomenology for different ratios $M_h/M_e <0$, demonstrating that exciton dissociation does not require fine-tuning: it is guaranteed to happen at large enough couplings. On the other hand, the exciton is always stable if $M_h/M_e >0$ (see panel (d) of Fig. \ref{fig:ldos_coupling_diags}),  because in this case the cloud created by the exciton is larger than the sum of the individual clouds created by the two unbound carriers, further stabilizing the exciton.\cite{burovski08,gerlach90}

\subsection{Exciton dissociation phase diagram}
\begin{figure}[t]
\centering
\includegraphics[width=0.48\textwidth]{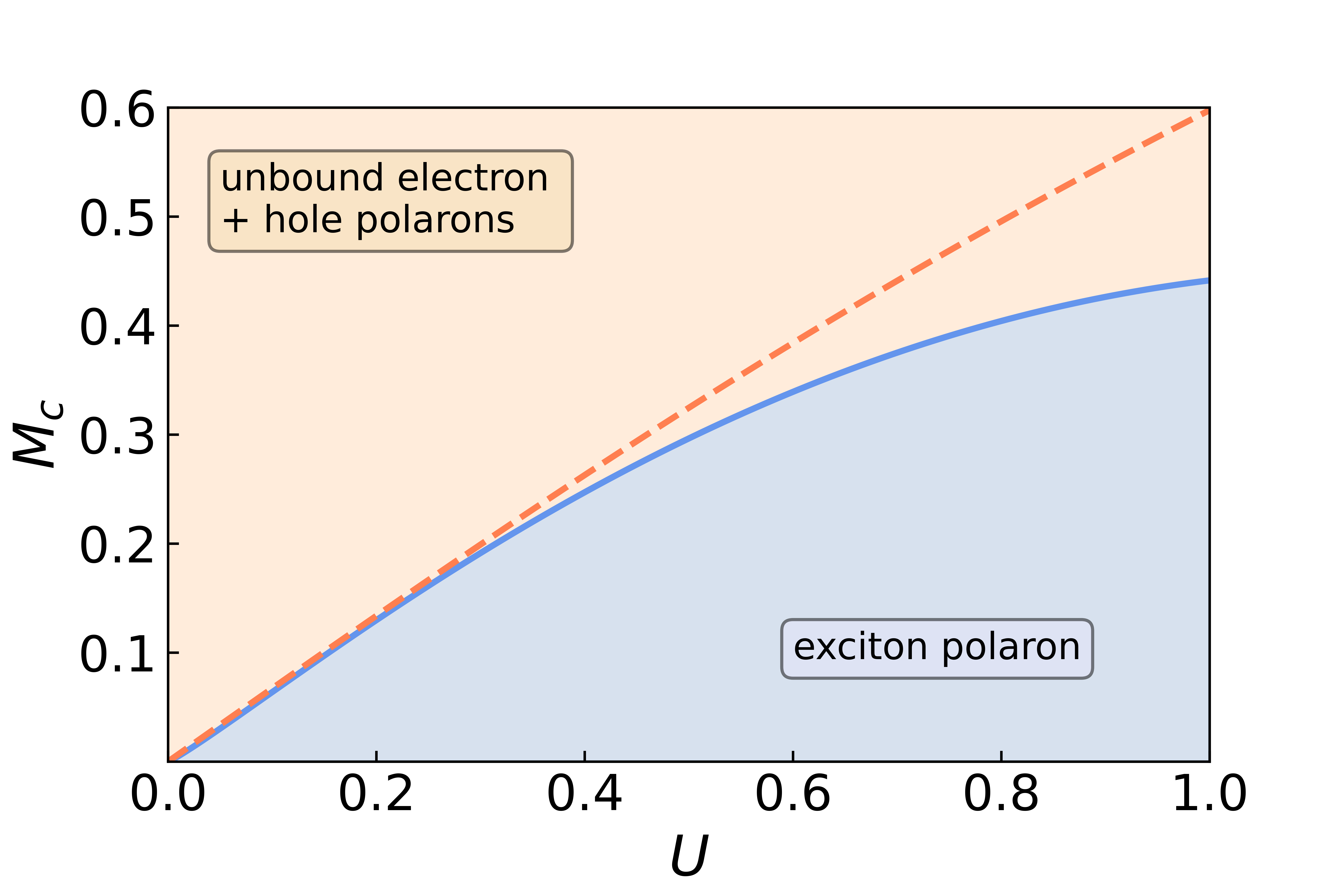}
				\caption{Exciton dissociation phase diagram. The critical electron-phonon coupling for dissociation increases with the Coulomb attraction $U$: it is calculated with MA (blue solid) and with perturbation theory (orange dashed). The orange region above the critical line indicates the region where we expect dissociated electron and hole polarons, whereas the blue region below the line represents the bound exciton-polaron region. Other parameters are $\Omega=0.5$, $M_e=-M_h$.
				}
		\label{fig:phase_diag}
\end{figure}

Figure \ref{fig:phase_diag}  traces the crossover (blue line) between the ground-states with an exciton-polaron and those with unbound electron- and hole-polarons. The dashed line shows the perturbation theory prediction (details in Appendix \ref{app:pert_exc_diss}). The agreement is excellent at small $U$, as expected,  while at larger $U$ perturbation theory overestimates the critical coupling needed for dissociation.

\subsection{Exciton-polaron characteristics} Next, we calculate the average number of phonons $N_{\text{ph}}$ in the exciton cloud, and also the probability $\rho_{ij}$ that the electron is at a distance $|j-i|$ from the hole in the exciton ground-state, scaled such that $\rho_{ii} =1$ (see Sec. \ref{sec:methods_char} for details).

\begin{figure}[t]
    \centering
    \includegraphics[width=0.9\columnwidth]{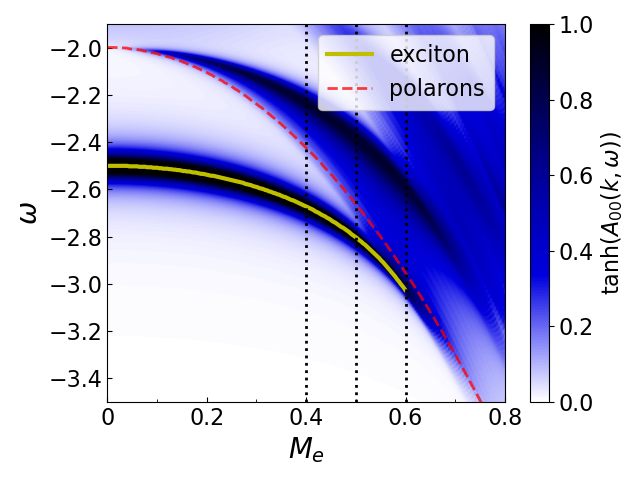}
    \includegraphics[width=0.72\columnwidth]{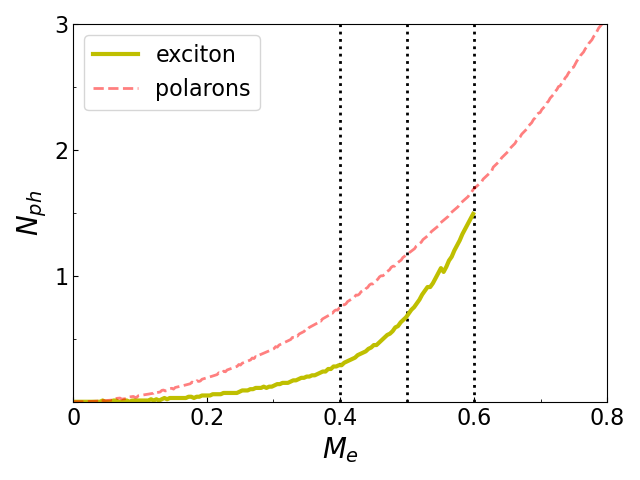}
    \includegraphics[width=0.3\columnwidth]{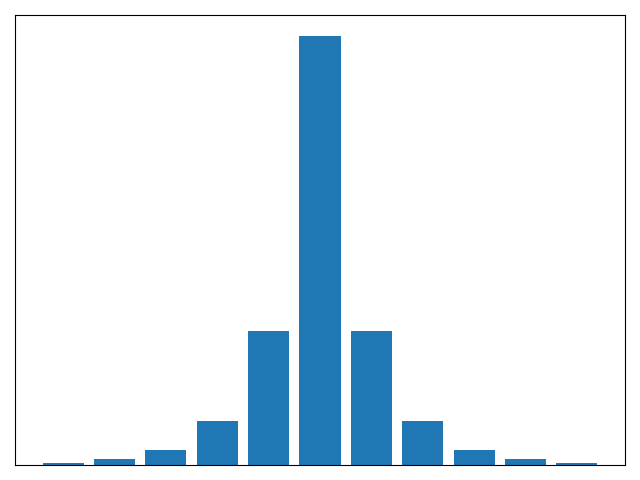}
    \includegraphics[width=0.3\columnwidth]{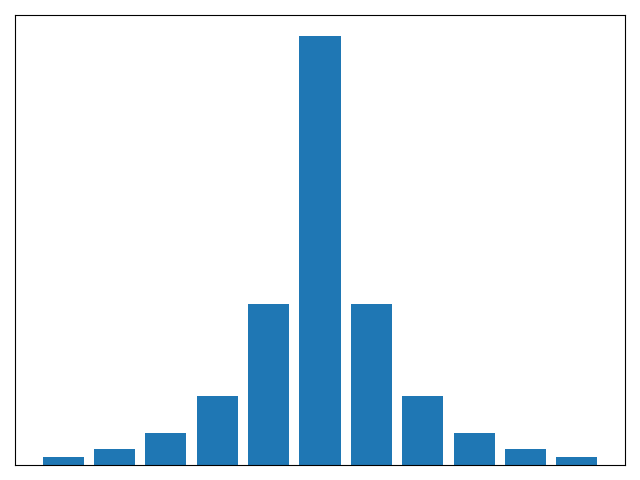}
    \includegraphics[width=0.3\columnwidth]{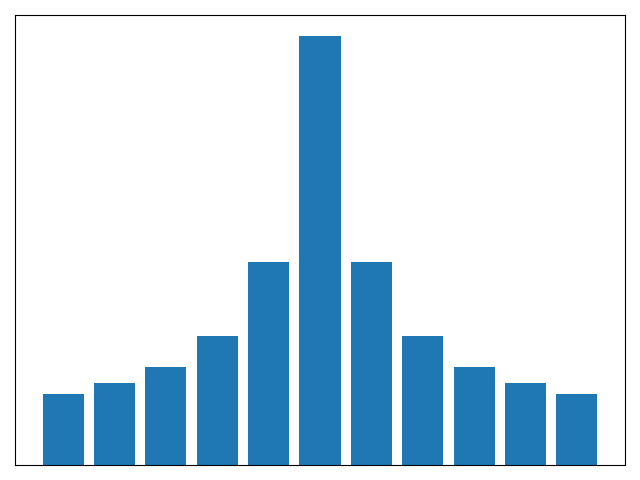}
    \caption{Characterization of the phonon cloud of the exciton-polaron. a) Contour plot of the LDOS at the hole site, as a function of the coupling $M_e$ and energy $\omega$. The yellow solid line tracks the exciton energy while the dashed red line tracks the lower edge of the continuum;  their intersection marks the dissociation point.  We track the exciton energy up to $M_e=0.6$, where its binding energy becomes comparable to $\eta$. b) Average number of phonons $N_{\text{ph}}$ in the exciton cloud (solid yellow line) and in the combined electron- and hole-polaron clouds (red dashed line).  c)-e) Probability $\rho_{ij}$ that the electron is at a distance $|j-i|$ from the hole in the exciton ground-state, scaled such that $\rho_{ii} =1$, for $M_e=0.4, 0.5, 
    0.6,$ respectively. Other parameters are  $\Omega = 0.5$, $U=1.5$, $M_e=-M_h$, $\eta=0.01$, $n_m = k_m = 20$, $|l-i|_m = 50$. }
    \label{fig:phcloud}
\end{figure}

Representative results are shown in Fig. \ref{fig:phcloud}. For completeness, panel (a) shows the LDOS versus $\omega$
and $M_e$, with dissociation occurring slightly above $M_e=0.6$.
Panel (b) shows $N_{\text{ph}}$ of the exciton-polaron (solid yellow line), compared to the sum of the ground-state average numbers of phonons in the electron-polaron and the hole-polaron clouds (red dashed line); the latter are calculated individually and then summed. As expected,  when tightly bound by an attractive $U$, the electron and the hole largely cancel each other's lattice distortions, resulting in many fewer phonons than for the free polarons. 

Panels (c)-(e) show $\rho_{ij}$ {\it vs.} $j-i$ for  $M_e = 0.4, 0.5, 0.6$, respectively (see vertical dotted lines in panels (a) and (b)). At small couplings, $\rho_{ij}$ is sharply peaked at the hole site $i$, as expected for a strongly bound, small  Frenkel exciton.  As the coupling increases,  $\rho_{ij}$ acquires ``fat tails'', that are consistent with a larger exciton. Just before dissociation, $\rho_{ij}$ spreads over very many sites, consistent with the smooth crossover  to an unbound electron-polaron that is (nearly) equally likely to be at any distance from the hole. 

\section{Conclusions}
\label{sec:discussion}
We have shown that strong carrier-phonon coupling favors the dissociation of excitons into free polarons, even on  1D chains where excitons should be stable for  any electron-hole attraction.  This phenomenology is the counterpart to what drives BCS superconductivity.\cite{bardeen57} There, phonons overscreen the electron-electron repulsion turning it into an effective attraction. Here, phonons screen the electron-hole attraction and can turn it repulsive, at sufficiently strong coupling.

This phenomenology is robust and should be considered when analyzing exciton stability in materials with carrier-phonon coupling because the critical coupling for  dissociation need not be very large. Figure \ref{fig:phcloud} shows a critical value $M_e=-M_h\approx 0.6$, which corresponds to a weak effective Holstein coupling  $\lambda_c = M_e^2/2t\Omega \approx 0.36$ for the electron, even though the bare exciton binding energy is a considerable $0.5t$ for those parameters. Indeed,  panel (b) of Fig. \ref{fig:phcloud} confirms that the average phonon numbers are small. Of course, to some extent this is because of the rather large phonon frequency $\Omega=0.5 t$ used there, although such ratios are reasonable in some organic materials.

Regarding the main approximations in our model: 


(i) we do not expect different dimensionality to change this phenomenology. In 3D, a bare exciton is stable only if the Coulomb attraction is above a critical value.\cite{burovski08}  Whether the critical value is 0 (like in 1D) or finite (like in 3D) is irrelevant:  strong enough carrier-phonon coupling will lower the effective attraction below this critical value and make the exciton unstable. Our  MA method can be straightforwardly used to study higher-D systems.

(ii) the assumption that the hole is immobile is also not essential: `releasing' the hole does not change this picture qualitatively, only quantitatively. Moreover, in the context of OSC materials doped with either acceptor or donor molecules, it is possible to envision trapping one species of the carriers on such molecules.

(iii) the assumptions that the coupling is to a single optical mode and that it is of Holstein type are also not essential. Regardless of such details, polaron formation associated with local excess charge leads to a lowering of the energy. That is the only ingredient necessary for the mechanism discussed here.

\begin{figure}[t]
    \centering
    \includegraphics[width=0.5\columnwidth]{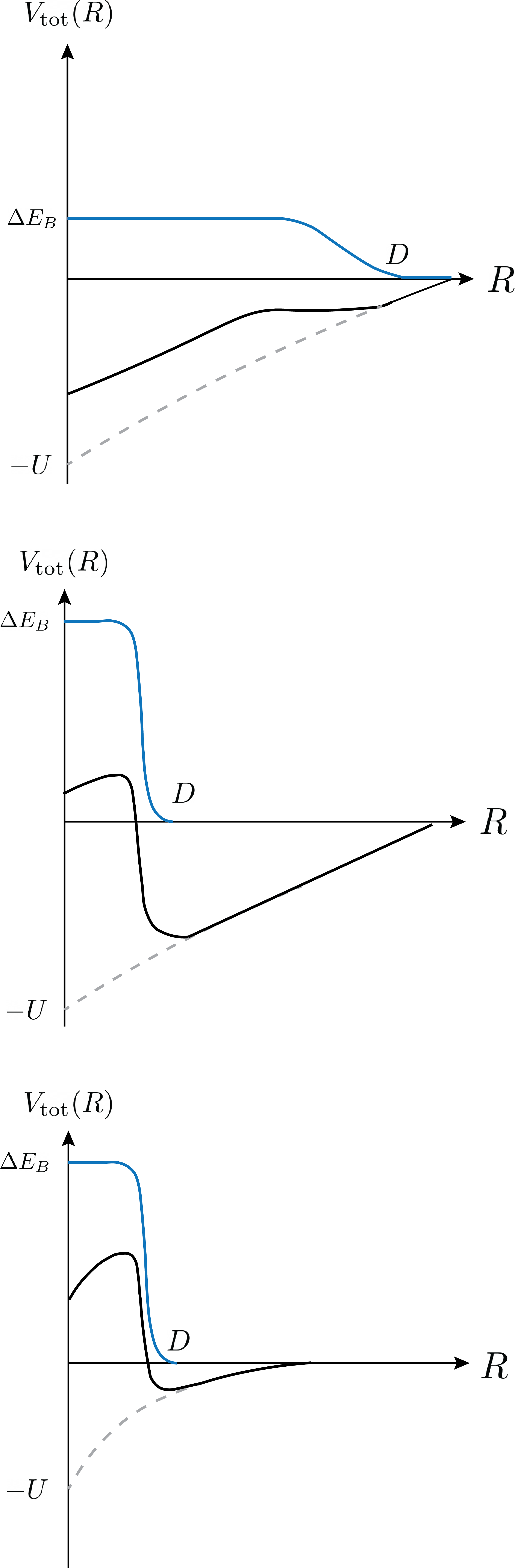}
    \caption{Schematic of the effective potential for the exciton-polaron. Screened electron-hole interaction (black line), obtained by summing the bare long-range Coulomb attraction (dashed line) and the contribution from phonon screening (blue line). Top: when the coupling is weak, the combined polaron radius $D$ is large and the screening is weak. Middle: strong coupling leads to small polarons with a strong short-range repulsion. The total potential has a minimum at $r \sim D$. Bottom: for a rapidly-decreasing bare attraction, a metastable exciton may be trapped at the $r = 0$ local minimum, before tunneling into a dissociated state.}
    \label{fig:potential}
\end{figure}

(iv) The assumption of a short-range Coulomb attraction is non-trivial, however, and relaxing it can lead to qualitative changes. This is because the phonon screening discussed here acts only at electron-hole distances $r< D$, where $D$ is the sum of the radii of the two polarons. If the electron and hole are sufficiently far, so that each can create its polaron cloud ($r > D$), the phonon screening vanishes. This contribution looks roughly like the blue lines in Fig. \ref{fig:potential}, where $\Delta E_B \approx - 2 M_eM_h/\Omega$ is the difference between the exciton-polaron and the free polarons formation energies. While $\Delta E_B$ increases with increasing coupling, $D$ decreases as the polarons become smaller. If the Coulomb attraction decreases rather slowly with $r$ (dashed line), it is possible that as the coupling goes from weak (top panel) to strong (middle panel), the total potential has a well whose minimum moves from $r\sim 0$  to $r\sim D$. The latter well can still trap a stable exciton in 1D, because both the lower dimensionality and the increased effective mass of strongly-coupled  Holstein polarons would favor a bound state. We believe that this explains why exciton dissociation was not observed in Ref. \onlinecite{hohenadler07}. However, in higher dimensions relevant for OSCs and/or  for lighter Peierls polarons,\cite{marchand10} such a `donut'-shaped trap might not suffice to bind the polarons and the ground-state at strong coupling would still exhibit dissociation.

A new scenario can occur if  the bare Coulomb attraction decreases significantly from $r=0$ to $r=D$. As sketched in Fig. \ref{fig:potential}(c), $r=0$  can be a local minimum of the total potential (black line) followed by a potential barrier and a very shallow potential well for $r > D$. A Frenkel exciton with radius smaller than $D$ can then be  metastable, with a lifetime inversely proportional to the probability of tunneling through the barrier.

Even though the ground state is the dissociated state,  small excitons  loaded optically into the metastable state might live long enough to control the OSC's behavior.
This may explain the very puzzling fact that some OSC materials, like pure C$_{60}$ films, have {\em both} very strongly bound excitons\cite{ishijima96,schlaich95} {\em and} finite, albeit small, charge separation efficiency. \cite{Zhang:2011} The latter would represent the small fraction of excitons that tunnel out and dissociate. This scenario is also qualitatively consistent 
with the observation that a dilute ($\sim$10\%) concentration of donor molecules increases the charge separation efficiency. Such molecules boost light absorption, so the metastable exciton state is populated more efficiently. This will increase the concentration of charge-separated pairs accordingly if the donor molecules are dilute enough to allow charge separation to proceed, explaining why peak efficiency occurs at a very low donor molecules concentration. \cite{Zhang:2011}
The above scenario cannot be verified with MA; however, a recent study found a  weak potential barrier due to nonlocal phonon screening in lead halide perovskites.\cite{park22} While their parameters are very different than ours, their finding supports the possible appearance of this new scenario in the right circumstances.

The results presented in this work illustrate some of the interesting physics expected in the many OSCs that have strong carrier-phonon coupling, and point towards possible ways to exploit it. We plan to investigate some of these topics in more detail in future works.

\acknowledgments
We thank Sarah Burke for bringing this problem to our attention and for many useful discussions. We thank David Reichman, Holger Fehske and Krzysztof Bieniasz for insightful comments. We acknowledge support from the Max Planck-UBC-UTokyo Centre for Quantum Materials and the Canada First Research Excellence Fund, Quantum Materials and Future Technologies Program of the Stewart Blusson Quantum Matter Institute, and from the Natural Sciences and Engineering Research Council of Canada (NSERC). We gratefully acknowledge the use of computing resources from the Stewart Blusson Quantum Matter Institute computing cluster LISA.

\appendix 

\section{Free carrier Green's function} 
\label{app:impurity_Greens_funs}

The Green's function $G_{ij}^{(i,0)}(z)$ corresponding to  $\hat{H}_0 = \hat{T}_e +\hat{V}_{e-h} +\hat{H}_{ph}$, \textit{i.e.} for the system without carrier-phonon coupling, can be calculated analytically. In the absence of electron-phonon coupling there is only an electron hopping on a 1D tight-binding lattice, subject to an on-site attractive potential from the static hole located at $i$. The corresponding Hamiltonian is $H_0=T - U c_i^{\dagger}c_i $. Here we calculate its lattice Green's function:
\begin{equation}
G^{(i,0)}_{lj}(z)=\bra{0}c_l[z-H_0]^{-1}c_j^{\dagger}\ket{0}
\end{equation}
Applying Dyson's identity, we find the EOM:
\begin{equation}
G^{(i,0)}_{l,j}(z)=g_{l-j}(z) - U g_{i-j}(z)G^{(i,0)}_{l,i}(z)
\label{eq:S1}
\end{equation}
where the free lattice Green's functions $g_{l-j}(z)=\bra{0}c_l[z-T]^{-1}c_j^{\dagger}\ket{0}$
can be calculated analytically: $g_\delta(z)=|\zeta(z)|^{|\delta|}/\sqrt{z-2t} \sqrt{z+2t}$, with $\zeta(z) = {z/2t} - \sqrt{z/2t-1}\sqrt{z/2t +1}$. 

Equation (\ref{eq:S1}) can be solved trivially to find:
\begin{equation}
G^{(i,0)}_{li}(z) = G^{(i,0)}_{l-i}(z)= \frac{g_{l-i}(z)}{1+Ug_0(z)} 
\end{equation}
and 
\begin{equation}
G^{(i,0)}_{ll}(z)=g_{0}(z) - U \frac{[g_{i-l}(z)]^2}{1+Ug_0(z)}.
\end{equation}

The propagators $\tilde{G}^{(i,0)}_{il}(z)$ appearing in the main text and in other appendices have the same expressions but with $U \rightarrow \tilde{U}$, where $\tilde{U}$ is the overscreened Coulomb attraction defined in Sec. \ref{sec:methods}.

\section{Green's function with carrier-lattice coupling}
\label{app:inter_Greens_funs}

Here the MA equations of motion are obtained by repeated application of the Dyson identity $\tilde{G}(\tilde{z})=\hat{G}^{(i)}_0(\tilde{z})+ \tilde{G}(\tilde{z})\hat{V}_{e-ph}\hat{G}^{(i)}_0(\tilde{z})$ where $\hat{G}^{(i)}_0(z)=(z-\hat{T}_e+\tilde{U}\hc{c}_i c_i)^{-1}$ is the resolvent in the absence of electron-phonon coupling. We note that its corresponding Green's functions $\tilde{G}^{(i,0)}_{ij}(z) = \langle 0 | c_i \hat{G}^{(i)}_0(z) c^\dagger_j |0\rangle$
equal those calculated in Sec. \ref{app:impurity_Greens_funs} upon replacing $U \rightarrow \tilde{U}$.

Using Dyson's identity once, we find:
\begin{multline}
H_{ij}(n,\tilde{z}) = \tilde{G}^{(i,0)}_{ij}(\tilde{z}-n\Omega) \left(\frac{M_h}{\Omega}\right)^n e^{-M_h^2/2\Omega^2} + \\
+ \tilde{G}^{(i,0)}_{ij}(\tilde{z}-n\Omega)M_e \left[nH_{ii}(n-1,\tilde{z})+H_{ii}(n+1,\tilde{z})\right]+
\\
 +\sum_{l\neq i}\tilde{G}^{(i,0)}_{lj}(\tilde{z}-n\Omega) M_e F_{ill}(n,1,\tilde{z}).
\label{eq:H_ij_expand_dyson}
\end{multline}
Here, the terms on the 2nd line arise when the electron travels to site $i$ and adds to or removes from the phonons already present there, while the last line describes terms where the electron moves to some other site $l$ and starts a new cloud there, with the corresponding generalized two-cloud propagator:
\begin{equation}
    F_{ijl}(n,k,\tilde{z}) \equiv \bra{0} c_ih_i \mathcal{U}_i\tilde{G}(\tilde{z}) \hc{h}_i \hc{c}_j  (\hc{b}_i)^n (\hc{b}_l)^k \ket{0}.
    \label{eq:f_def}
\end{equation}
The equation of motion (\ref{eq:H_ij_expand_dyson}) is  exact.  Solving it necessitates calculating the  propagators $F_{ill}$ that appear in it. We generate their equations of motion using again the Dyson identity, but now also imposing the variational constraint consistent with the one-site MA$^{(0)}$ approximation for the electron cloud, namely that additional phonons cannot be created away from the two existing clouds. 
The resulting EOMs are
\begin{widetext}
\begin{align}
F_{ill}(n,k,\tilde{z}) =& M_e\tilde{G}^{(i,0)}_{ll}(\tilde{z}-(n+k)\Omega)\left[k F_{ill}(n,k-1,\tilde{z}) +F_{ill}(n,k+1,\tilde{z})\right] \nonumber\\
&+M_e\tilde{G}^{(i,0)}_{il}(\tilde{z}-(n+k)\Omega)\left[k F_{iil}(n,k-1,\tilde{z}) +F_{iil}(n,k+1,\tilde{z})\right] \nonumber\\
F_{iil}(n,k,\tilde{z}) =& M_e\tilde{G}^{(i,0)}_{ii}(\tilde{z}-(n+k)\Omega)\left[k F_{iil}(n,k-1,\tilde{z}) +F_{iil}(n,k+1,\tilde{z})\right] \nonumber\\
&+M_e\tilde{G}^{(i,0)}_{il}(\tilde{z}-(n+k)\Omega)\left[k F_{iil}(n,k-1,\tilde{z}) +F_{iil}(n,k+1,\tilde{z})\right].
\label{eq:f_system}
\end{align}
\end{widetext}


Eqs. (\ref{eq:H_ij_expand_dyson}-\ref{eq:f_system}) define a linear, inhomogeneous system of coupled equations that can be numerically solved for each value of $z$, with the resulting $H_{ij}(n,\tilde{z})$ then used in Eq. (\ref{eq:coup_exc_prop}) to construct $G_{ij}(z)$. However, this approach is computationally intensive because one needs large cutoffs for the maximum numbers $k_{m}, n_{m}$ of phonons in the two clouds, as well as for the maximum distance $|l-i|_m$ between the clouds, before convergence is reached. An improved approach is discussed in Appendix \ref{app:simplify}.

\section{Simplifying the EOMs}
\label{app:simplify}
A much more efficient yet still accurate solution to Eqs. (\ref{eq:H_ij_expand_dyson}-\ref{eq:f_system}) can be obtained by taking advantage of the fact that for the energies of interest, which lie below the free electron continuum, the free propagators $\tilde{G}^{(i,0)}_{il}(z)$ decrease exponentially with the distance $|l-i|$. If we keep only the largest term with $l=i$, then Eqs. (\ref{eq:f_system}) split into two uncoupled recurrence relations, one for $F_{ill}$ and one for $F_{iil}$, with only the former needed in Eq.  (\ref{eq:H_ij_expand_dyson}). This former recurrence relation can be solved with the ansatz:
\begin{equation}
F_{ill}(n,k,\tilde{z}) = A^{(i,l)}_k(\tilde{z}-n\Omega)F_{ill}(n,k-1,\tilde{z})
\end{equation}
where we note that $F_{ill}(n,0,\tilde{z})\equiv H_{il}(n,\tilde{z})$. The continued fractions
\begin{equation}
A^{(i,l)}_k(z) = 
\frac{kM_e  \tilde{G}^{(i,0)}_{ll}(z-k\Omega)}{1-M_e \tilde{G}^{(i,0)}_{ll}(z-k\Omega)A^{(i,l)}_{k+1}(z)}
\end{equation}
are calculated starting from $A^{(i,l)}_{k_{m}+1}(z)=0$ for a sufficiently large $k_{m}$ to ensure the desired accuracy. In particular, this means that we can replace $F_{ill}(n,1,\tilde{z}) = A^{(i,l)}_1(\tilde{z}-n\Omega)H_{il}(n,\tilde{z})$ in Eq. (\ref{eq:H_ij_expand_dyson}) to convert it into a linear system linking only the $H_{ij}$ propagators. This still requires a summation over all the sites in the system, which in practice means summing over sites $l$ up to a distance large enough from $i$ that the sum converges. 

An efficient solution of such a linear system was proposed in Refs. \onlinecite{MonaEPL89(2010),HadiPRB85(2012)} and we adopt it here. It is based on the observation that for $|l-i| \gg 1$, the local potential $\tilde{U}$ created by the hole  becomes irrelevant and the impurity Green's function reduces to the free electron propagator 
\begin{equation}
    \tilde{G}_{ll}^{(i,0)}(\tilde{z}) \rightarrow g_0(\tilde{z})= \frac{1}{N}\sum_k \frac{1}{\tilde{z}-\epsilon_k} = \frac{1}{\sqrt{\tilde{z}-2t}\sqrt{\tilde{z}+2t}}.
\end{equation} 
As a result, for $|l-i|\gg 1$, the continued fractions approach an asymptotic value that becomes independent of $i,l$:  $A_1^{(i,l)}(\tilde{z}-n\Omega)\rightarrow \Sigma_{MA}(\tilde{z}-n\Omega)/M_e$ . Physically, $\Sigma_{MA}(z)$ is the MA$^{(0)}$ self-energy of the electron-polaron in the absence of the `impurity' potential created by the hole located at $i$ (see Ref. \onlinecite{berciu06} for a derivation)
\begin{equation}
 \Sigma_{MA}(z)
 = \frac{M_e^2 g_0(z-\Omega)}{1-\cfrac{2M_e^2  g_0(z-\Omega)g_0(z-2\Omega)}{1- \cfrac{3M_e^2  g_0(z-2\Omega)g_0(z-3\Omega)}{1- \dots}}}
 \label{sigMA}
\end{equation}

Because this asymptotic value is independent of $l$, we can define a renormalized energy
\begin{equation}
v_{il}(\tilde{z}-n\Omega) = M_e A_1^{(i,l)}(\tilde{z}-n\Omega) - \Sigma_{MA}(\tilde{z}-n\Omega)
\end{equation}
which vanishes fast with increasing $|l-i|$. The sum in Eq. (\ref{eq:H_ij_expand_dyson}) can be recast in terms of it by renormalizing the energy argument of the free propagators:
\begin{widetext}
\begin{align}
H_{ij}(n,\tilde{z}) = \tilde{G}^{(i,0)}_{ij}(\tilde{\tilde{z}}_n) \left(\frac{M_h}{\Omega}\right)^n e^{-\frac{M_h^2}{2\Omega^2}} +& \tilde{G}^{(i,0)}_{ij}(\tilde{\tilde{z}}_n) M_e \left[nH_{ii}(n-1,\tilde{z})+H_{ii}(n+1,\tilde{z})\right] \nonumber\\
& +\sum_{l\ne i}\tilde{G}^{(i,0)}_{lj}(\tilde{\tilde{z}}_n) v_{il}(\tilde{z}-n\Omega) H_{il}(n,\tilde{z})
\label{eq:H_ij_approx_final}
\end{align}
\end{widetext}
where we defined $\tilde{\tilde{z}}_n \equiv \tilde{z}-n\Omega-\Sigma_{MA}(\tilde{z}-n\Omega). $
Equations  (\ref{eq:H_ij_approx_final}) converge much more quickly with the summation over $l$ and can be solved efficiently.

The accuracy of the approximation of replacing the coupled Eqs. (\ref{eq:H_ij_expand_dyson}-\ref{eq:f_system}) with the much more compact and efficienct Eq. (\ref{eq:H_ij_approx_final}) is validated in Appendix
\ref{app:approx_full_compare}.
 
\begin{figure}[t]
    \centering
    \includegraphics[width=0.48\columnwidth]{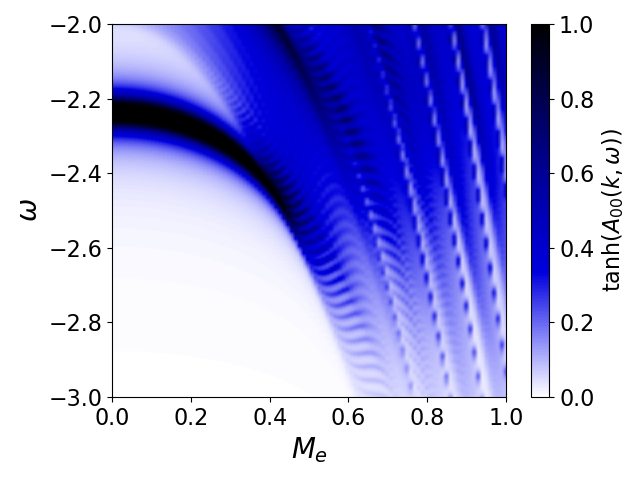}
    \includegraphics[width=0.48\columnwidth]{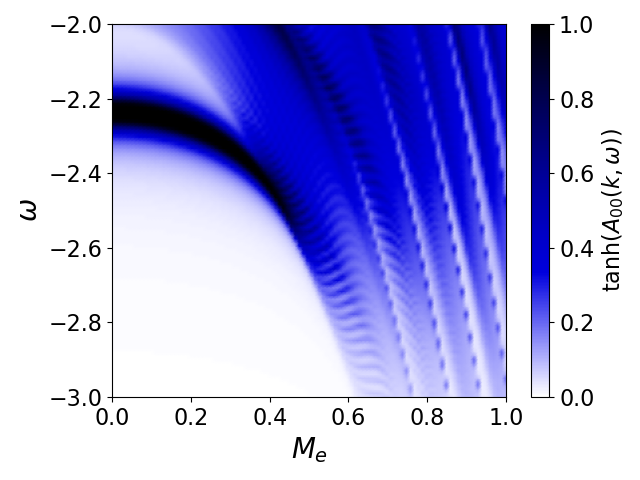}
\caption{Comparison of the LDOS at the hole site  from solving the full variational solution described by  Eqs. (\ref{eq:coup_exc_prop}),(\ref{eq:H_ij_expand_dyson}),(\ref{eq:f_system}), shown in the left panel, versus the simplified and much more efficient Eq. (\ref{eq:H_ij_approx_final}), shown in the right panel. Visually, the two are nearly indistinguishable, with most differences coming in at higher energies. Model parameters are $U = 1, \Omega = 0.5, \eta = 0.01, M_e = -M_h$ and convergence parameters are $n_m = k_m = 12, |l-i|_m = 50$. }
\label{fig:approx_vs_exact}
\end{figure}

\section{Full vs approximate variational solutions}
\label{app:approx_full_compare}
The full variational solution of the particle+hole propagator can be obtained by simultaneously solving Eqs. (\ref{eq:coup_exc_prop}), (\ref{eq:H_ij_expand_dyson}) and (\ref{eq:f_system}). They can be solved numerically, but this is slow because exceedingly large truncation cutoffs (system sizes) are required for convergence.
Above in Appendix \ref{app:simplify}, we proposed a much more efficient approximation which replaces Eqs. (\ref{eq:H_ij_expand_dyson})-(\ref{eq:f_system}) with Eqs. (\ref{eq:H_ij_approx_final}).

To validate this approximation, in Fig. \ref{fig:approx_vs_exact} we show a typical comparison of the results of the two methods for the LDOS at the hole site, focusing on the lower-energy part of the spectrum, of interest for the dissociation issue. Evidently, the agreement is very good. Similar diagrams were produced in all parameter regimes explored in this paper, thus effectively validating our approximation.

\section{Perturbation theory for exciton dissociation}
\label{app:pert_exc_diss}

Here we summarize the perturbation theory (PT) calculation used to draw the dissociation line in Fig. \ref{fig:phase_diag} in the main text. We begin by estimating the ground-state energies for the individual polarons. The result for the (static) hole-polaron is $E_P^h=-M_h^2/\Omega$. To find the electron-polaron's PT counterpart, we use the single polaron Green's function at the same one-site MA$^{(0)}$ level of approximation:\cite{berciu06}
$
G(k,z) = [z-\epsilon_{k}-\Sigma_{MA}(z)]^{-1}
$
where the full expression for $\Sigma_{MA}(z)$ is shown in Eq. (\ref{sigMA}). To lowest non-trivial order in PT, it becomes  $\Sigma_{MA} \approx M_e^2 g_0(\omega-\Omega)$. Using this expression to find the lowest $k=0$ pole, we find the polaron ground-state energy to be:
\begin{equation}
E^e_{P}(k=0) = -2t-\frac{M_e^2}{\sqrt{\Omega(\Omega+4t)}}
\end{equation}
The PT-predicted lower edge of the continuum is then at $E_{min}=E^e_{P}(k=0)+E_P^h$.

To find the bound exciton energy, we proceed similarly, essentially solving the EOMs to lowest order in the couplings, and then finding the location of the lowest peak for $k=0$. For simplicity, we only list here the result when $M_e=-M_h$. We find the exciton ground-state energy to be given by $E_{exc}=z_0+\alpha g_0(z_0-\Omega) M_e^2$ where $z_0 = -\sqrt{U^2+4t^2} $ is the bare exciton energy, and 
\begin{equation}
\alpha = \frac{4 G^{(i,0)}_{ii}(z_0)[g_0(z_0-\Omega)-2 G^{(i,0)}_{ii}(z_0-\Omega)]}{1+2 G^{(i,0)}_{ii}(z_0)[g_0(z_0-\Omega)- 2G^{(i,0)}_{ii}(z_0-\Omega)] }
\nonumber
\end{equation}

The dissociation occurs when $E_{exc}=E_{min}$.


%

\end{document}